\documentstyle[aps,pra,epsf]{revtex}

\begin{document}
\draft

\newcommand{\nc}{\newcommand}
\newcommand{\BE}{\begin{equation}}
\newcommand{\EE}{\end{equation}}
\newcommand{\BA}{\begin{eqnarray}}
\newcommand{\EA}{\end{eqnarray}}

\title{Estimation of B\"uttiker-Landauer traversal time\\
based on the visibility of transmission current}

\author{
Koh'ichiro Hara$^{1}$\footnote{E-mail khara@hep.phys.waseda.ac.jp},
and Ichiro Ohba$^{1,2,3}$\footnote{E-mail ohba@mn.waseda.ac.jp}
}

\address{
$^1$Department of Physics, Waseda University Tokyo 169-8555, Japan,\\
$^2$Kagami Memorial Laboratory for Materials Science and Technology,\\
Waseda University Tokyo 169-0051, Japan,\\
$^3$Advanced Research Center for Science and Technology\\
Waseda University, Tokyo 169-8555, Japan
}

\maketitle

\begin{abstract}

We present a proposal for the estimation of B\"uttiker-Landauer
traversal time based on the visibility of transmission current.   We
analyze the tunneling phenomena with a time-dependent potential and
obtain the time-dependent transmission current.  We found that the
visibility was directly connected to the traversal time.  Furthermore,
this result is valid not only for rectangular potential barrier but also
for general form of potential to which the WKB approximation is
applicable .  We compared these results with the numerical values
obtained from the simulation of Nelson's quantum mechanics.  Both of
them fit together and it shows our method is very effective to measure
experimentally the traversal time.

\end{abstract}


\pacs{PACS number(s): 03.65.Bz, 73.40.Gk}

\section{Introduction}
Soon after the advent of quantum mechanics, MacColl suggested that there is a time associated with the passage of a particle under a tunneling barrier, i.e. a tunneling time \cite{MacColl}.  Now the time has been measured in several experiments and its qualitative results have been obtained.  However, it is not clear whether a unique time exists or not, since we have no univocal definition of tunneling time and no definite experimental data.   See \cite{Hauge=1989}, \cite{Landauer=1994} and references therein for reviews of the problem.

In this paper, we present a proposal for the estimation of B\"uttiker-Landauer traversal time based on the visibility of transmission current experimentally.  B\"uttiker and Landauer\cite{Buttiker=1982}, \cite{Buttiker=1983} invoked an oscillatory barrier to estimate a tunneling time.  The original static barrier was augmented by a small oscillation in the barrier height.   The amplitude of the oscillation is kept small; the disturbance of the original
kinetics can be made small as desired.   At very low modulation frequencies the incident particle sees a particular part of the modulation cycle.   The particle sees an effectively static barrier, but later parts of the incident wave see a slightly different barrier height.   As one turns up the modulation frequency, one eventually reaches a range where an incident particle no longer sees a particular portion of the modulation cycle, but is affected by a substantial part of the modulation cycle, or several cycles.   They claimed that the frequency at which this transition occurs, i.e., the frequency where one begins to deviate substantially from the adiabatic approximation, is an indication of the length of time that a particle interacts with the barrier.   They made carefully several comments as follows: It is, of course, an approximate indication of a time scale.   It is not the eigenvalue of a Hamiltonian, indicative of a precisely measurable value.   Moreover, this traversal time value may really be characteristic of a statistical distribution.

They showed that for an opaque rectangular barrier, the modulated
barrier approach yields
\BE
\tau = dm/\hbar \kappa \ ,
\label{tra t}
\EE
where $d$ is the barrier length and $\hbar\kappa$ the magnitude of the
imaginary momentum under the barrier.   For a potential that allows the
WKB approximation, it yields
\BE
\tau = \int_{B}dx\frac{m}{\hbar\kappa(x)}\ ,
\label{WKBtra t}
\EE
where $B$ means the barrier region.

This gives a plausible estimation of traversal time based on a
theoretical background.   However, if one wants to measure the value of
traversal time by an experiment, one has to draw it from the asymptotic
behavior of transmission rate as a function of $\omega$.  Generally its
dependence on $\omega$ does not change so rapidly, that one cannot
easily estimate the value from experimental data.  There is another type
of experiment; one projects a stationary incident particle beam on the
target with oscillating barrier and measure the time dependence of
transmission current which may also oscillate with the same frequency.  
Here we show the visibility of oscillating current gives us a good
information about traversal time.

\section{Time-dependent barrier}
Following \cite{Buttiker=1982}, \cite{Buttiker=1983} and \cite{Martin=1993}, we start by considering a Hamiltonian,
\BE
H=-\frac{\hbar^2}{2m}\frac{d^2 \ }{dx^2} + V_{0}(x) + V_{1}(x)\cos
\omega t,
\label{totalH}
\EE
where $V_{0}(x)$ is static and $V_{1}(x)$ is the amplitude of a small
modulation.  Incident particles with energy $E$ interacting with the
perturbation $V_{1}\cos\, \omega t$, will emit or absorb modulation
quanta $\hbar\omega$. The Schr\"odinger equation of this Hamiltonian has
the solution in the barrier region
\BE
\Psi(x,t;E')=\phi_{E'}(x)\exp\left(
-i\frac{E't}{\hbar}\right)\sum_{n=-\infty}^{n=\infty}J_{n}\left(\frac{V_{1}}{\hbar\omega}\right){\rm
e}^{-in\omega t},
\label{Psi}
\EE
where $\phi_{E'}(x)$ is an eigenfunction of the time-independent
Hamiltonian $H_{0}= - (\hbar^2 /2m)d^2 /dx^2 + V_{0},\ 
H_{0}\phi_{E'}=E'\phi_{E'}$ and $J_{n}$ is a Bessel function.  
The time modulation of the potential gives rise to sidebands describing
 particles which have absorbed ($n>0$) or emitted ($n<0$) modulation 
quanta. 
Therefore we have to take into account the many sidebands of which the
Bessel functions are appreciable.

To the left of the barrier, we allow an incident wave at energy $E$ and
reflected waves at energies $E'=E_{n}\equiv E + n\hbar\omega$,
\BE
\Psi^{\rm I}(x,t)= {\rm e}^{ikx}\,{\rm
e}^{-i\frac{E}{\hbar}t}+\sum_{E_{n}>0}A_{n}{\rm e}^{-ik_{n}x}{\rm
e}^{-i\frac{E_{n}}{\hbar}t}, 
\label{PsiI}
\EE
where $k_{n}=\sqrt{\frac{2mE_{n}}{\hbar^2}}$ , $E_{0}=E$ and $k_{0}=k$. 
See Fig.1.   We consider only the positive energy solutions.  In the
barrier region, in addition to the solution (\ref{Psi}) with $E'=E$,
there exist other evanescent (and oscillating, in a certain case) modes
corresponding to the reflected wave with energy $E'=E_{n}$. Here we also
consider only positive energy solutions.  Taking account of these
points, we have a solution in the barrier region,
\BE
\Psi^{\rm I{}I}(x,t)=\sum_{E_{n}>0}^{n_{\rm eff}}{\rm
e}^{-i\frac{E_{n}}{\hbar}t}
\sum_{m}^{n_{\rm eff}}\left(B_{m}{\rm e}^{\kappa_{m} x} +C_{m}{\rm e}^{-\kappa_{m}
x}\right)J_{n-m}\left(\frac{V_{1}}{\hbar\omega}\right),
\label{PsiII}
\EE
where $\kappa_{n} =\sqrt{\frac{2m(V_{0}-E_{n})}{\hbar^2}}$.  For the
transmitted wave, we have
\BE
\Psi^{\rm I{}I{}I}(x,t)=\sum_{E_{n}>0}D_{n}{\rm e}^{ik_{n}x}{\rm
e}^{-i\frac{E_{n}}{\hbar}t} .
\label{PsiIII}
\EE
For small $V_{1}$, $J_{n}$ is proportional to $(V_{1}/2\hbar\omega)^n$
and thus, only the small numbers of terms in the summation of
(\ref{PsiII}) contribute effectively.  Correspondingly the numbers of terms in the summations of (\ref{PsiI}) and (\ref{PsiIII}) are suppressed.   To find the solution for the Schr\"odinger equation, we match a superposition of incident and
reflected waves (\ref{PsiI}), and also transmitted waves (\ref{PsiIII}),
at each energy $E_{n}$, to solutions within the barber (\ref{PsiII}).
 As a result of somewhat tedious but straight calculation(see the Appendix A), 
we have the transmission and reflection coefficients in the leading order,
\begin{eqnarray}
D_{n} & = & \frac{J_{n}(V_{1}/\hbar\omega)}{J_{0}(V_{1}/\hbar\omega)}\,
\frac{2D_{0}{\rm e}^{i(k-k_{n})d/2}}{\det(k_{n},\kappa_{n})}\nonumber\\
 \quad & & \times \left\{ (\kappa_{n}^2 - k_{n}k_{0})\sinh\kappa_{n}d 
 - (\kappa^2 -
k_{n}k_{0})(\kappa_{n}/\kappa_{0})\sinh\kappa_{0}d\right.\nonumber\\
 \quad & & \qquad + \left. i\kappa_{n}(k_{n}+k_{0})(\cosh\kappa_{0}d -
\cosh\kappa_{n}d)\right\},
 \end{eqnarray}
and
\begin{eqnarray}
A_{n} & = & \frac{J_{n}(V_{1}/\hbar\omega)}{J_{0}(V_{1}/\hbar\omega)}\,
\frac{D_{0}{\rm e}^{i(k-k_{n})d/2}}{\det(k_{n},\kappa_{n})}\nonumber\\
 \quad & & \times \left\{ (\kappa_{n}^2 - k_{n}k_{0})\sinh\kappa_{n}d
\cosh\kappa_{0}d  \right.\nonumber\\
 \quad & & \quad - (\kappa^2 +
k_{n}k_{0})(\kappa_{n}/\kappa_{0})\cosh\kappa_{n}d\sinh\kappa_{0}d
\nonumber\\
 \quad & & \quad + 
i\kappa_{n}(k_{0}-k_{n})(1-\cosh\kappa_{n}d\cosh\kappa_{0}d)\nonumber\\
 \quad & & \quad \left.
-i((k_{0}\kappa_{n}^2/\kappa_{0})-k_{n}\kappa_{0})\sinh\kappa_{n}d\sinh\kappa_{0}d
\right\},
 \end{eqnarray}
where $\det (k_{n},\kappa_{n})$ is defined by
\begin{eqnarray}
\det (k_{n},\kappa_{n}) & \equiv & \left|
\begin{array}{cc}
(\kappa_{n} + ik_{n}){\rm e}^{-\kappa_{n}d} & -(\kappa_{n} -
ik_{n})\nonumber\\
(\kappa_{n} - ik_{n}){\rm e}^{\kappa_{n}d} & -(\kappa_{n} + ik_{n})
\end{array}
\right|\\
\quad & = & 2(\kappa_{n}^2 - k^2 _{n})\sinh\,\kappa_{n} d
-4ik_{n}\kappa_{n}\cosh\,\kappa_{n} d.
\end{eqnarray}

From these results we can obtain the transmission probability defined by
the ratio of transmitted current $j_{\rm I{}I{}I}$ and the incident
current $j_{\rm inc}=\hbar k/m$.   It depends on the time as well as the
position of measurement due to the interference among different energies
waves.   However, if we take a time average of the ratio, its dependence
will disappear,
\BE
{\bar T}= \sum_{n=0}^{n_{\rm eff}}\frac{k_n}{k_{0}}|D_n|^2 .
\EE
We show an example of numerical result of the time-averaged transmission
probability in Fig. 2.

Now we will discuss the traversal time.   As following to B\"uttiker and
Landauer, we assume that $\hbar\omega \ll E$, so that the wave numbers
of the sidebands are
\BE
k_{\pm n}=\sqrt{\frac{2m(E\pm \hbar\omega)}{\hbar}}\approx k\pm
n\frac{m\omega}{\hbar k},
\EE
and assume $\hbar\omega \ll V_{0} -E$, so that 
\BE
\kappa_{\pm n}=\sqrt{\frac{2m(V_{0}-E \mp \hbar\omega)}{\hbar}}\approx
\kappa\mp n\frac{m\omega}{\hbar \kappa}.
\EE
In the case of opaque barrier, taking account of the asymptotic forms
of transmitted wave amplitudes,
\BE
D_{\pm 1}=\pm\frac{V_{1}}{2\hbar \omega}D_{0} ({\rm
e}^{\pm\omega\tau}-1){\rm e}^{\mp i\frac{\omega\tau}{2}},
\label{asympt}
\EE
B\"uttiker and Landauer included first order corrections to the static
barrier and obtained the intensity for the transmitted sidebands, for
the case of small $V_{1}$,
\BE
T_{\pm 1}=\frac{k_{\pm 1}}{k_{0}}\left(\frac{V_{1}}{2\hbar
\omega}\right)^2 ({\rm e}^{\pm\omega\tau}-1)^2 T_{0},
\EE
where $\tau = md/\hbar\kappa$.  From this expression they found that
there exists the crossover from the low frequency behavior
\BE
T_{\pm 1} 
 = \frac{k_{\pm 1}}{k_{0}}\left(\frac{V_{1}\tau}{2\hbar}\right)^2 T_{0},
\EE
where the two intensities of the sidebands are equal, to the high
frequency behavior
\BE
T_{+1}=\frac{k_{+1}}{k_{0}}\left(\frac{V_{1}}{2\hbar \omega}\right)^2
{\rm e}^{2\omega\tau} T_{0},
\EE
\BE
T_{-1}=\frac{k_{-1}}{k_{0}}\left(\frac{V_{1}}{2\hbar \omega}\right)^2
T_{0},
\EE
where the two intensities differ strongly.   This transition to
imbalance is best described by 
\BE
\frac{k_{-1}T_{+1} - k_{+1}T_{-1}}{k_{-1}T_{+1} + k_{+1}T_{-1}} = \tanh
\omega\tau.
\EE
Thus they claimed the crossover from the low frequency behavior to the
high frequency behavior yields the traversal time.



\section{Visibility and traversal time}
Their claim is a very interesting idea to estimate a certain kind of
tunneling time, but it is rather difficult to determine its value from
experiments.   Now let us consider the time dependence of the
transmitted currents.  If one observes the currents at a fixed point
$x=L$, one may see the interference effect between the different
frequency waves in the first order approximation,
\begin{eqnarray}
T & = & \frac{1}{k_{0}}{\rm Re}\left\{\left( k_{0}D_{0}{\rm
e}^{i(k_{0}L-E_{0}t)} +  k_{1}D_{1}{\rm e}^{i(k_{1}L-E_{1}t)} + 
k_{-1}D_{-1}{\rm e}^{i(k_{-1}L-E_{-1}t)}  \right)^{*} \right.\nonumber\\
 & & \qquad\qquad\qquad\times \left. \left( D_{0}{\rm e}^{i(k_{0}L-E_{0}t)} + D_{1}{\rm e}^{i(k_{1}L-E_{1}t)} +  D_{-1}{\rm e}^{i(k_{-1}L-E_{-1}t)} 
\right) \right\} \nonumber\\
 & \sim & |D_{0}|^2 + \frac{1}{k_{0}}|D_{0}|\left((k_{0}+k_{+1})|D_{+1}|
+ (k_{0}+k_{-1})|D_{-1}|\right) \cos (\omega t - \phi(L)),
\end{eqnarray}
where
$\phi(L)$ is a phase which is independent on time $t$,
\begin{eqnarray}
\phi(L) & = & \phi_{+1}(L) = {\rm arg}\left(\frac{D_{+1}}{D_0}\right) + 
(k_{+1}-k_{0})L \nonumber\\
  & = & -\phi_{-1}(L) = -\left({\rm arg}\left( \frac{D_{-1}}{D_0}\right) + 
(k_{-1}-k_{0})L\right).
\end{eqnarray}
Here the asymptotic forms (\ref{asympt}) were used.  Now we show the
numerical result of the time dependence of transmitted current at a
fixed point in Fig.3.   
If a detector has a good time resolution, one may measure
this visibility of the transmitted wave
\begin{eqnarray}
I_{\rm vis} & \equiv & \frac{T_{\rm max}-T_{\rm min}}{T_{\rm max}+T_{\rm
min}}\nonumber\\
& = & \frac{1}{k_{0}}\left((k_{0}+k_{+1})\left|\frac{D_{+1}}{D_{0}}\right| +
(k_{0}+k_{-1})\left|\frac{D_{-1}}{D_{0}}\right|\right).\label{visi-1}
\end{eqnarray}
In the case of a small perturbation $V_{1}$ and an opaque static
potential, equation (\ref{visi-1}) is approximated by
\BE
I_{\rm vis} \sim \frac{2V_{1}}{\hbar\omega}\sinh\omega\tau, \label{vis}
\EE
from which the traversal time is expressed by the visibility as follows;
\BE
\tau = \frac{1}{\omega} \sinh^{-1}\left(
\frac{\hbar\omega}{2V_{1}}I_{\rm vis}\right).\label{tau-1}
\EE
If one can choose an experimental setup satisfying the condition
$\omega\tau \ll 1$, this expression becomes to
\BE
\tau \sim \frac{\hbar}{2V_{1}}I_{\rm vis}.\label{tau-2}
\EE

For the case of general potential shown in Fig.4, which allows the WKB
approximation, we have a transmitting wave after the potential wall, 
\begin{eqnarray}
\Psi^{\rm I{}I{}I}(x,t)
& = &
i\frac{4 S_{0}}{4+S_{0}^2}\frac{1}{\sqrt{k_0(x)}}
\exp \left\{i(\int_{x_{2}}^{x}k_0(x')dx'
- \frac{\pi}{4} ) \right\}{\rm e}^{-iEt/\hbar}\nonumber\\
\ & \times &
\left[1+\frac{J_{1}\left( V_{1}/\hbar\omega \right)}
{J_{0}\left( V_{1}/\hbar\omega \right)}
\sqrt{\frac{k_{0}(x)}{k_1(x)}}
\exp\left\{ i \int_{x_2}^{x}  \frac{m \omega}{\hbar k_0(x')} dx' \right\}
(1-\Sigma_{1}){\rm e}^{-i \omega t} \right. \nonumber\\
\ &+& \left. \frac{J_{-1}\left( V_{1}/\hbar\omega \right)}
{J_{0}\left( V_{1}/\hbar\omega \right)}
\sqrt{\frac{k_{0}(x)}{k_{-1}(x)}}
\exp\left\{ -i \int_{x_2}^{x}  \frac{m \omega}{\hbar k_0(x')} dx' \right\}
(1-\Sigma_{-1}){\rm e}^{i \omega t} \right], 
\end{eqnarray}
where
\begin{eqnarray}
\Sigma_{\pm 1}
& = &
\frac{4+S_{0}^2 }{4+S_{\pm 1}^2}
\ \frac{S_{\pm 1}}{S_{0}},\\
S_{n}
& = &
\exp \left(- \int_{x_{1}}^{x_{2}}\kappa_{n}(x)dx \right).
\end{eqnarray}
The detailed calculation is given in Appendix B.  For an opaque potential,
the damping factors $S_{n}$ are so small, that the transmitted current
becomes
\BE
T \sim S_{0}^{2}
\left\{1+2\frac{J_{1}\left(V_{1}/\hbar\omega \right)}
{J_{0}\left(V_{1}/\hbar\omega \right)}
(\Sigma_{-1} - \Sigma_{1})\cos \left(\omega t - \phi(x) \right) \right\},
\EE
where
\BE
\phi(x) = \int^{x}_{x_2}\frac{m \omega}{\hbar k_0(x')}dx'
\EE
Therefore the visibility is given by
\begin{eqnarray}
I_{\rm vis}
& = &
2\frac{J_{1}\left(V_{1}/\hbar\omega \right)}
{J_{0}\left(V_{1}/\hbar\omega \right)}
(\Sigma_{-1} - \Sigma_{1})\nonumber\\
& \sim &
\frac{V_{1}}{\hbar\omega}
2\sinh\left(\frac{m\omega}{\hbar}
\int_{x_{1}}^{x_{2}}\frac{1}{\kappa_0(x)}dx \right),\label{tau-3}
\end{eqnarray}
and Eq. (\ref{tau-2}) is replaced by the following expression,
\BE
\tau_{\rm WKB} = 
\frac{m}{\hbar}\int_{x_{1}}^{x_{2}}\frac{1}{\kappa_0(x)}dx \sim
\frac{\hbar}{2V_{1}}I_{\rm vis}.\label{tau-4}
\EE




\section{Comparison of numerical results with the simulation\\
based on the Nelson's quantum mechanics}
Here we evaluate the tunneling time by the use of Nelson's approach of
quantum mechanics \cite{Nelson=1966} and compare them with numerical
results of traversal time obtained from the visibility.   Nelson's
quantum mechanics, using the real-time stochastic process, enables us to
describe individual experimental runs of a quantum system in terminology
of the ``analog" of classical mechanics, i.e., the ensemble of sample
paths. These sample paths are generated by the stochastic process,
\BE
dx(t)=(u(x(t),t)+v(x(t),t)) dt + dw(t),\label{Langevin}
\EE
where $x(t)$ is a stochastic variable corresponding to the coordinate of
the particle, and $u(x(t),t)$ and $v(x(t),t)$ are the osmotic velocity
and the current velocity, respectively.  The $dw(t)$ is the Gaussian
white noise with the statistical properties of 
\BE
\langle dw(t) \rangle=0, 
\mbox{\quad and \quad}
\langle dw(t)dw(t) \rangle =\frac{\hbar}{m}dt.
\EE
In principle the osmotic and the current velocities are given by solving
coupled two equations, i.e., the kinetic equation and the
``Newton-Nelson equation".  The whole ensemble of sample paths gives us
the same results as quantum mechanics in the ordinary approach.  Once
the equivalence of Nelson's framework and ordinary quantum mechanics is proved, it is convenient to use the relation
\BE
u={\rm Re}\frac{\hbar}{m}\frac{\partial\ }{\partial x}\ln\,
\psi(x,t),\mbox{\quad and \quad}
v={\rm Im}\frac{\hbar}{m}\frac{\partial\ }{\partial x}\ln\, \psi(x,t),
\label{verocities}
\EE
where $\psi$ is the solution of Schr\"odinger equation. Since
individual sample path has its own history, we obtain information on the
time parameter, e.g., the traversal time \cite{Imafuku=1995}, \cite{Imafuku=1997}.

Now using the Nelson's quantum mechanics, we estimate the traversal time
crossing over a time-dependent potential barrier shown in Fig.1.  
Suppose a simulation of tunneling phenomena based on (\ref{Langevin}),
starting from $t=-\infty$ and ending at $t=\infty$. As we treat a wave
packet satisfying the time-dependent Schr\"odinger equation, the wave
packet is located in region I initially and turns finally into two
spatially separated wave packets which are in regions I and I{}I{}I.   
Fig.5 shows a typical transmission sample path calculated by Eq.(\ref{Langevin}) with ``backward time evolution method" \cite{Imafuku=1995}, \cite{Imafuku=1997}.  The traversal time using this approach, $\tau_{\rm Nelson}$, is defined as the averaged time interval in which the random variable $x(t)$ stays in the barrier region I{}I. 
Thus $\tau_{\rm Nelson}$ defined in this way has a character of
statistical distribution as pointed in \cite{Buttiker=1982}, \cite{Buttiker=1983}, since it is the value averaged over the ensemble of sample paths having the transmitting wave packets.

We call the traversal time obtained by the visibility of transmission current, 
$\tau_{\rm vis}$. Let us compare $\tau_{\rm vis}$ with 
$\tau_{\rm Nelson}$ and $\tau_{\rm WKB}$ in a rectangular potential 
barrier numerically.  
Here we take the unit with $m=\hbar=1$. Fig.6 shows these numerical 
results versus potential width $d$ and Fig.7 shows
those versus $V_0/E_0$, where $E_0$ is an incident energy and $V_0$ is 
a potential height.    It has been shown that, in the opaque case, the Fokker-Planck equation for the distribution for the samples can be solved analytically and gives $\tau_{\rm Nelson} \sim \frac{md}{\hbar \kappa}\left(=\tau_{\rm WKB}\right)$\cite{Imafuku=1995}, \cite{Imafuku=1997}.    The parameters adopted in Fig. 6 give an imaginary wave number $\kappa = 1$ in the unite of $k_0$ which corresponds to the opaque potential except for very thin potential barrier.   Thus we can see that $\tau_{\rm Nelson}$ and $\tau_{\rm WKB}$ agree with each other.   It is notable that $\tau_{\rm vis}$ fits also well with them except for thin barrier where the opaqueness condition is broken.   The imaginary wave number dependence of traversal time is shown in Fig. 7 for a fixed and rather thick potential barrier width.   The value of $\kappa$ becomes larger than 1 for $V_{0} > 2E_{0}$ and in this region $\tau_{\rm Nelson}$ agrees with $\tau_{\rm WKB}$.   On the other hand, in the region $V_{0} < 2E_{0}$, $\tau_{\rm WKB}$ becomes to deviate from $\tau_{\rm Nelson}$, where the opaqueness condition is not satisfied.   However $\tau_{\rm vis}$ can reproduce the value of $\tau_{\rm Nelson}$ for almost all region.  From these two figures, we see, in the opaque case, that $\tau_{\rm Nelson}$ coincide with $\tau_{\rm WKB}$ with respect to its dependence on potential width $d$ and on the imaginary wave number $\kappa$.   While there is an obvious reason why the $\tau_{\rm WKB}$ can only applicable to the opaque case, one needs not assume any approximation to evaluate $\tau_{\rm Nelson}$ in principle.  Therefore the latter may represent an characteristic property of time scale for tunneling phenomena not only for the opaque case but also for the translucent case.   However, both of these traversal times are defined only on the bases of theoretical models, but cannot be checked by experiment so easily.  It should be noticed that $\tau_{\rm vis}$ is connected to the experimental data directly, and the theoretical estimation may be checked by experiment rather easily.  Thus we think that $\tau_{\rm vis}$ can be a good candidate presenting time scale of tunneling phenomena both for the opaque case and for the translucent case.



\section{Summary and comments}
In this paper, we present a proposal for the estimation of
B\"uttiker-Landauer traversal time based on the visibility of
transmission current.   We analyzed the tunneling phenomena with a
time-dependent potential described by Eq. (\ref{totalH}), and obtained the
time-dependent transmission current for a small perturbation $V_{1}$ and
an opaque case.    We found that the visibility is directly connected to
the traversal time, while B\"uttiker and Landauer proposed that the
crossover from the low frequency behavior to the high frequency
behavior yields the traversal time.  Furthermore, this result is valid
not only for rectangular potential barrier but also for general form of
potential to which the WKB approximation is applicable.   After a brief
review of Nelson's quantum mechanics, by which the traversal time is
calculated definitely, we compared those results with the numerical
values obtained from the simulation of Nelson's framework.  Both of them
fit together not only for the opaque case but also for the translucent case and it shows our method is very effective to measure
experimentally the traversal time.

\section{Acknowledgment}
K. Imafuku joined with us in the early stage of this work.  The authors
acknowledge his interest and also useful and helpful discussion with K.
Imafuku, H. Nakazato and Y. Yamanaka.



\appendix
\section{Amplitudes}

In this Appendix, we recapitulate briefly how to determine the amplitudes of the sidebands at $E \pm n \hbar \omega$ from the matching conditions \cite{Buttiker=1982}, \cite{Buttiker=1983}, \cite{Martin=1993}.   It is convenient to define the following quantities:
\begin{eqnarray}
W^{(n)}_{m}(x) & \equiv & (B_{m}{\rm e}^{\kappa x}+C_{m}{\rm e}^{-\kappa
x})
J_{n-m}\left(\frac{V_{1}}{\hbar\omega}\right),\\
W^{(n)'}_{m}(x)& \equiv & \kappa_{m}(B_{m}{\rm e}^{\kappa x}-C_{m}{\rm
e}^{-\kappa x})J_{n-m}\left(\frac{V_{1}}{\hbar\omega}\right),
\end{eqnarray}
where the prime means a derivative with respect to the coordinate $x$.   At the energy $E_{n}$, we have the matching conditions,
\begin{eqnarray}
\delta_{n0}\,{\rm e}^{-i\alpha_{n}} + A_{n}\,{\rm e}^{i\alpha_{n}} & = & 
\sum_{m}W^{(n)}_{m}\left( -\frac{d}{2}\right),\label{matching1}\\
i\frac{2\alpha_{n}}{d}\left( \delta_{n0}\,{\rm e}^{-i\alpha_{n}} -
A_{n}\,{\rm e}^{i\alpha_{n}}  \right)& = & 
\sum_{m}W^{(n)'}_{m}\left( -\frac{d}{2}\right),\label{matching2}\\
D_{n}\,{\rm e}^{i\alpha_{n}} & = & 
\sum_{m}W^{(n)}_{m}\left( \frac{d}{2}\right),\label{matching3}\\
i\frac{2\alpha_{n}}{d}D_{n}\,{\rm e}^{i\alpha_{n}} & = & 
\sum_{m}W^{(n)'}_{m}\left( \frac{d}{2}\right),\label{matching4}
\end{eqnarray}
where $\alpha_{n}\equiv k_{n}d/2$.   Noticing that $J_{n}$ is
proportional to $(V_{1}/2\hbar\omega)^n$ and taking only the leading
terms, we approximate equations (\ref{matching1}), (\ref{matching2}),
(\ref{matching3}) and (\ref{matching4}) and obtain the transmission
coefficients and reflection coefficients.  At the energy $E$, we recover
the results of static barrier, the reflection and the transmission
coefficients,
\begin{eqnarray}
A_{0} & = & \frac{-2(k^2 + \kappa^2 )\sinh\,\kappa d}{\det
(k,\kappa)}\,{\rm e}^{-ikd},\\
D_{0} & = & \frac{-4ik\kappa}{\det (k,\kappa)}\,{\rm e}^{-ikd},
\end{eqnarray}
where $\det (k,\kappa)$ is defined by
\begin{eqnarray}
\det (k,\kappa) & \equiv & \left|
\begin{array}{cc}
(\kappa + ik){\rm e}^{-\kappa d} & -(\kappa - ik)\nonumber\\
(\kappa - ik){\rm e}^{\kappa d} & -(\kappa + ik)
\end{array}
\right|\\
\quad & = & 2(\kappa^2 - k^2 )\sinh\,\kappa d -4ik\kappa\cosh\,\kappa d.
\end{eqnarray}
Similarly, at the energy $E_{n}$, we have the transmission and
reflection coefficients in the leading order,
\begin{eqnarray}
D_{n} & = & \frac{J_{n}(V_{1}/\hbar\omega)}{J_{0}(V_{1}/\hbar\omega)}\,
\frac{2D_{0}{\rm e}^{i(k-k_{n})d/2}}{\det(k_{n},\kappa_{n})}\nonumber\\
 \quad & & \times \left\{ (\kappa_{n}^2 - k_{n}k_{0})\sinh\kappa_{n}d 
 - (\kappa^2 -
k_{n}k_{0})(\kappa_{n}/\kappa_{0})\sinh\kappa_{0}d\right.\nonumber\\
 \quad & & \qquad + \left. i\kappa_{n}(k_{n}+k_{0})(\cosh\kappa_{0}d -
\cosh\kappa_{n}d)\right\},
 \end{eqnarray}
and
\begin{eqnarray}
A_{n} & = & \frac{J_{n}(V_{1}/\hbar\omega)}{J_{0}(V_{1}/\hbar\omega)}\,
\frac{D_{0}{\rm e}^{i(k-k_{n})d/2}}{\det(k_{n},\kappa_{n})}\nonumber\\
 \quad & & \times \left\{ (\kappa_{n}^2 - k_{n}k_{0})\sinh\kappa_{n}d
\cosh\kappa_{0}d  \right.\nonumber\\
 \quad & & \quad - (\kappa^2 +
k_{n}k_{0})(\kappa_{n}/\kappa_{0})\cosh\kappa_{n}d\sinh\kappa_{0}d
\nonumber\\
 \quad & & \quad + 
i\kappa_{n}(k_{0}-k_{n})(1-\cosh\kappa_{n}d\cosh\kappa_{0}d)\nonumber\\
 \quad & & \quad \left.
-i((k_{0}\kappa_{n}^2/\kappa_{0})-k_{n}\kappa_{0})\sinh\kappa_{n}d\sinh\kappa_{0}d
\right\},
 \end{eqnarray}
where $\det (k_{n},\kappa_{n})$ is defined by
\begin{eqnarray}
\det (k_{n},\kappa_{n}) & \equiv & \left|
\begin{array}{cc}
(\kappa_{n} + ik_{n}){\rm e}^{-\kappa_{n}d} & -(\kappa_{n} -
ik_{n})\nonumber\\
(\kappa_{n} - ik_{n}){\rm e}^{\kappa_{n}d} & -(\kappa_{n} + ik_{n})
\end{array}
\right|\\
\quad & = & 2(\kappa_{n}^2 - k^2 _{n})\sinh\,\kappa_{n} d
-4ik_{n}\kappa_{n}\cosh\,\kappa_{n} d.
\end{eqnarray}



\section{The visibility in a general potential case}

We give an expression for the visibility in a general potential case by
the use of the WKB approximation.  A stationary solution $\Psi_{E}(x)$
with energy $E$ satisfies the Schr\"odinger equation
\BE
\left[ -\frac{\hbar^2}{2m}\nabla^2 +V(x)
\right]\Psi_{E}(x)=E\Psi_{E}(x).
\EE
 Stating from the outgoing wave solution in the region I{}I{}I, 
\BE
\Psi^{\rm I{}I{}I}(x) =
\frac{1}{\sqrt{k(x)}}
\exp \left\{i \left(\int_{x_{2}}^{x}k(x')dx' 
- \frac{\pi}{4} \right) \right\},
\EE
we have the evanescent wave solution in the region I{}I,
\BE
\Psi^{\rm I{}I}(x) =
\frac{1}{\sqrt{\kappa(x)}}
\left[-\frac{i}{S}\exp \left\{-\int_{x_{1}}^{x}\kappa(x')dx' \right\}
+ \frac{S}{2}\exp \left\{\int_{x_{1}}^{x}\kappa(x')dx' \right\} \right],
\EE
and then, the incoming and reflecting wave solutions
\begin{eqnarray}
\Psi^{\rm I}(x) &=& \frac{1}{\sqrt{k(x)}}
\left[ \left(-i\frac{4+S^2}{4S}\right)
\exp \left\{-i \left(\int_{x}^{x_{1}}k(x')dx'-\frac{\pi}{4} \right)
 \right\} \right. \nonumber\\
& + &
\left. \left( -i\frac{4-S^2}{4S} \right)
\exp \left\{i \left(\int_{x}^{x_{1}}k(x')dx' -\frac{\pi}{4} \right) \right\}
 \right],
\end{eqnarray}
where
\BE
S=\exp \left( -\int_{x_{1}}^{x^{2}}\kappa(x')dx' \right).
\EE

Using these stationary WKB solutions, we can write down a time dependent
solution in the case shown in Fig.4,
\BE
\Psi(x,t)=\left\{
\begin{array}{ll}
\sum_{E} \Psi_{E}^{\rm I}(x) {\rm e}^{-iEt/\hbar} & \quad 
(x \le x_{1}),\\[2.5mm]
\sum_{E} \Psi_{E}^{\rm I{}I}(x)
\sum_{n}J_{n}\left(\frac{V_{1}}{\hbar\omega}\right)
{\rm e}^{-i(E+n\hbar\omega)t/\hbar}& \quad 
(x_{1} \le x \le x_{2}),\\[2.5mm]
\sum_{E} \Psi_{E}^{\rm I{}I{}I}(x) 
{\rm e}^{-iEt/\hbar} & \quad 
(x_{2} \le x).
\end{array}
\right.
\EE
For the case of $n=0$, we have the solution
\BE
\Psi_{E_{0}}(x)=\left\{
\begin{array}{l}

D_{0}J_{0}\left(\frac{V_{1}}{\hbar\omega}\right) 
\frac{1}{\sqrt{k(x)}} 
\left[ \left( -i\frac{4+S_{0}^2}{4S_{0}}\right)
\exp\{-i(\int_{x}^{x_{1}}k(x')dx'-\frac{\pi}{4} )\}\right. \\
 \qquad\quad  \left. + \left( -i\frac{4-S_{0}^2}{4S_{0}}\right)
\exp\{i(\int_{x}^{x_{1}}k(x')dx' -\frac{\pi}{4} )\}  \right],
  \quad (x\le x_{1}),\\[2.5mm]

D_{0}J_{0}\left(\frac{V_{1}}{\hbar\omega}\right)
\frac{1}{\sqrt{\kappa(x)}}
\left[-\frac{i}{S_{0}}\exp\{-\int_{x_{1}}^{x}\kappa(x')dx'\} \right. \\
\qquad\qquad\qquad\qquad \left. +
\frac{S_{0}}{2}\exp\{\int_{x_{1}}^{x}\kappa(x')dx'\} \right] \quad
(x_{1}\le x\le x_{2}),\\[2.5mm]

D_{0}J_{0}\left(\frac{V_{1}}{\hbar\omega}\right)
\frac{1}{\sqrt{k(x)}}\exp\{i(\int_{x_{2}}^{x}k(x')dx' -\frac{\pi}{4} )\}
\qquad (x_{2}\le x),

\end{array}
\right.
\EE
where $S_{n}$ is the damping factor of the $n$-th mode,
\BE
S_{n}=\exp\left(-\int_{x_{1}}^{x_{2}}\kappa_{n}(x)dx\right).
\EE
The coefficients are fixed by the incoming wave normalization,
\BE
D_{0}J_{0}\left(\frac{V_{1}}{\hbar\omega}\right)=
i\frac{4S_{0}}{4+S_{0}^{2}}.
\EE
For the case of $n=1$, we have to consider the two types of wave in the
region I{}I,
\begin{eqnarray}
\ & D_{0}J_{1} \left(\frac{V_{1}}{\hbar\omega}\right)
\frac{1}{\sqrt{\kappa(x)}}
\left[-\frac{i}{S_{0}}\exp\{-\int_{x_{1}}^{x}\kappa(x')dx'\} +
\frac{S_{0}}{2}\exp\{\int_{x_{1}}^{x}\kappa(x')dx'\} \right]\nonumber\\
\ & + D_{1}J_{0} \left(\frac{V_{1}}{\hbar\omega}\right)
\frac{1}{\sqrt{\kappa_{1}(x)}}
\left[-\frac{i}{S_{1}}\exp\{-\int_{x_{1}}^{x}\kappa_{1}(x')dx'\}
+\frac{S_{1}}{2}\exp\{\int_{x_{1}}^{x}\kappa_{1}(x')dx'\} \right],\nonumber\\
\end{eqnarray}
both of which should be matched to the reflecting and the transmitting
waves with energy $E_{1}$ and wave number $k_{1}$.  A similar relation
holds for $n=-1$.  Requiring the condition that there are no incoming
waves in these modes, we can determine the coefficients $D_{\pm 1}$ 
in the region I{}I{}I,
\BA
D_{\pm 1}J_{0}\left(\frac{V_{1}}{\hbar\omega}\right) & = &  
-\frac{4+S_{0}^2}{4+S_{\pm 1}^{2}} \ 
\frac{S_{\pm 1}}{S_{0}} \ D_{0}J_{\pm 1}
\left(\frac{V_{1}}{\hbar\omega}\right).
\EA
Considering the above results, we get the transmitting wave up to $n=\pm 1$,
\begin{eqnarray}
\Psi^{\rm I{}I{}I}(x,t)
& = &
i\frac{4 S_{0}}{4+S_{0}^2}\frac{1}{\sqrt{k_0(x)}}
\exp \left\{i(\int_{x_{2}}^{x}k_0(x')dx'
- \frac{\pi}{4} ) \right\}{\rm e}^{-iEt/\hbar}\nonumber\\
\ & \times &
\left[1+\frac{J_{1}\left( V_{1}/\hbar\omega \right)}
{J_{0}\left( V_{1}/\hbar\omega \right)}
\sqrt{\frac{k_{0}(x)}{k_1(x)}}
\exp\left\{ i \int_{x_2}^{x}  \frac{m \omega}{\hbar k_0(x')} dx' \right\}
(1-\Sigma_{1}){\rm e}^{-i \omega t} \right. \nonumber\\
\ &+& \left. \frac{J_{-1}\left( V_{1}/\hbar\omega \right)}
{J_{0}\left( V_{1}/\hbar\omega \right)}
\sqrt{\frac{k_{0}(x)}{k_{-1}(x)}}
\exp\left\{ -i \int_{x_2}^{x}  \frac{m \omega}{\hbar k_0(x')} dx' \right\}
(1-\Sigma_{-1}){\rm e}^{i \omega t} \right], 
\end{eqnarray}
where
\BE
\Sigma_{\pm 1}= \frac{4+S_{0}^2 }{4+S_{\pm 1}^2 }\, 
\frac{S_{\pm 1}}{S_{0}}.
\EE



\newpage

\begin{figure}
  \epsfxsize=9cm
  \centerline{\epsfbox{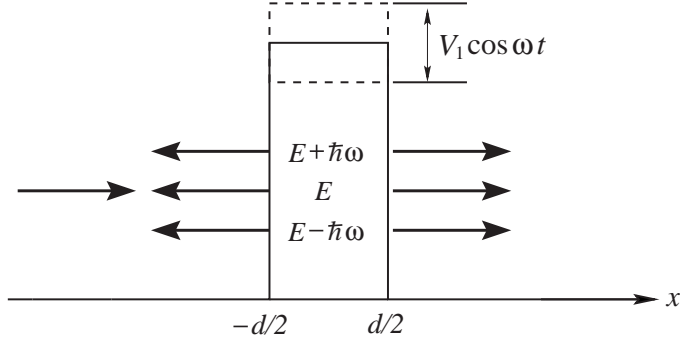}}
  \caption{Particles transmitted or reflected at a barrier of height $V_0$ and width $d$ interacting a small modulation $V_1 \cos \omega t$ can absorb or emit modulation quanta $\hbar \omega$. The transmitted and reflected waves contain amplitudes at the frequency $E/\hbar$ and the sideband frequencies $E_n/\hbar$.}
  \label{Fig.1}
\end{figure}

\begin{figure}
  \epsfxsize=9cm
  \centerline{\epsfbox{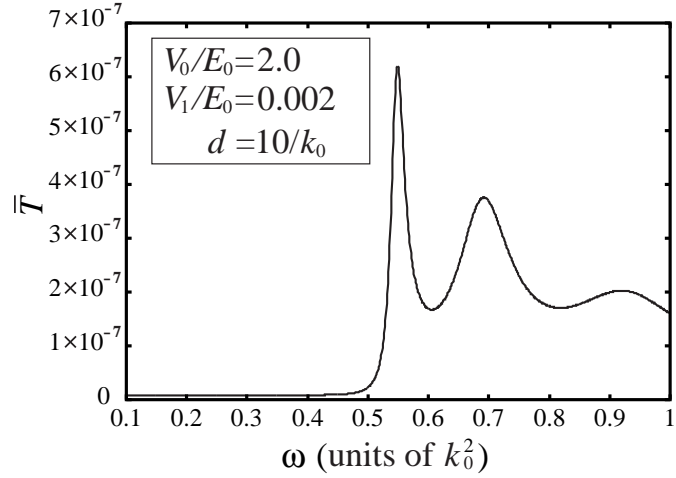}}
  \caption{The transmission probability taking a long time average.}
  \label{Fig.2}
\end{figure}

\begin{figure}
  \epsfxsize=10cm
  \centerline{\epsfbox{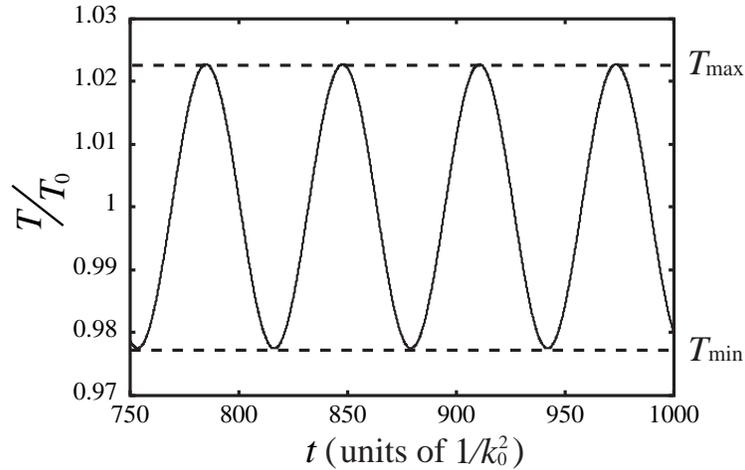}}
  \caption{The time dependence of the transmitted currents at a fixed point $x=750$(units of $1/k_0$). The potential frequency $\omega$ is $0.1$(units of $k_0^2$). Other parameters (static potential height, small modulation amplitude, etc.) are the same values in Fig.2. In this figure, $T_0$ is the transmission probability in the static potential case, that is, $V_1=0$.}
  \label{Fig.3}
\end{figure}

\begin{figure}
  \epsfxsize=9cm
  \centerline{\epsfbox{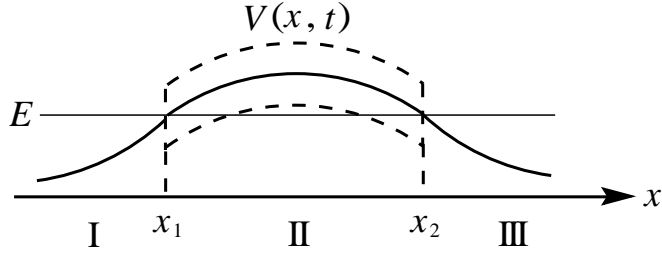}}
  \caption{Schematical illustration of one-dimensional tunneling in the general potential case. The small modulation $V_1 \cos \omega t$ exists in the region of II (illustrated with the dashed line).}
  \label{Fig.4}
\end{figure}

\begin{figure}
  \epsfxsize=9.5cm
  \centerline{\epsfbox{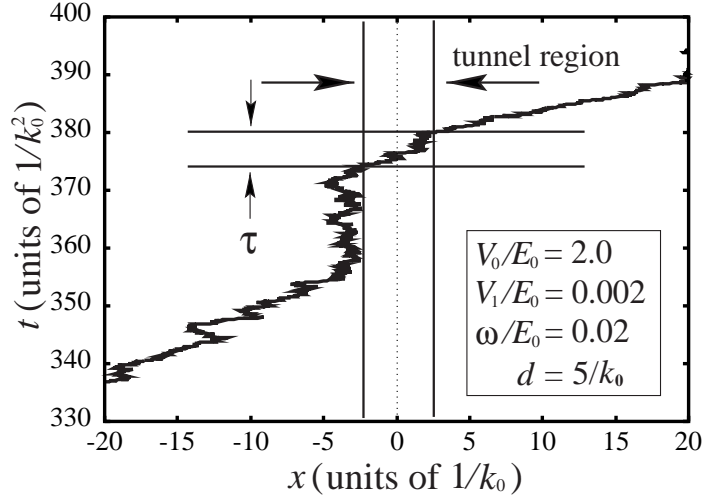}}
  \caption{Typical transmission sample path calculated by Eq.(\ref{Langevin}).}
  \label{Fig.5}
\end{figure}

\begin{figure}
  \epsfxsize=9cm
  \centerline{\epsfbox{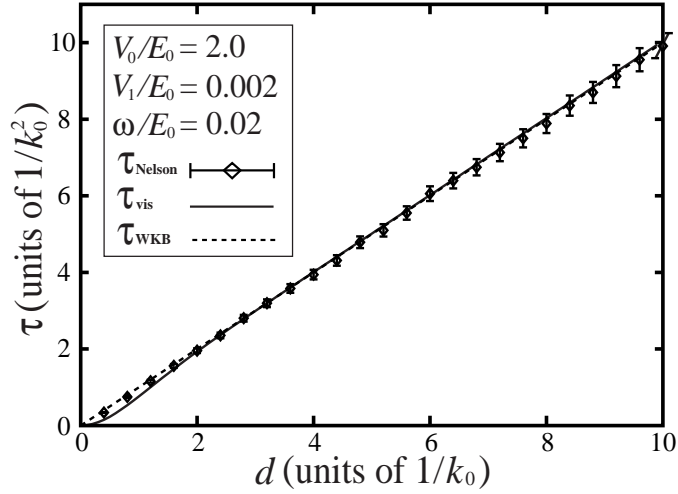}}
  \caption{Comparison of numerical results of traversal times versus potential width $d$ in a rectangular potential barrier.}
  \label{Fig.6}
\end{figure}

\begin{figure}
  \epsfxsize=9.25cm
  \centerline{\epsfbox{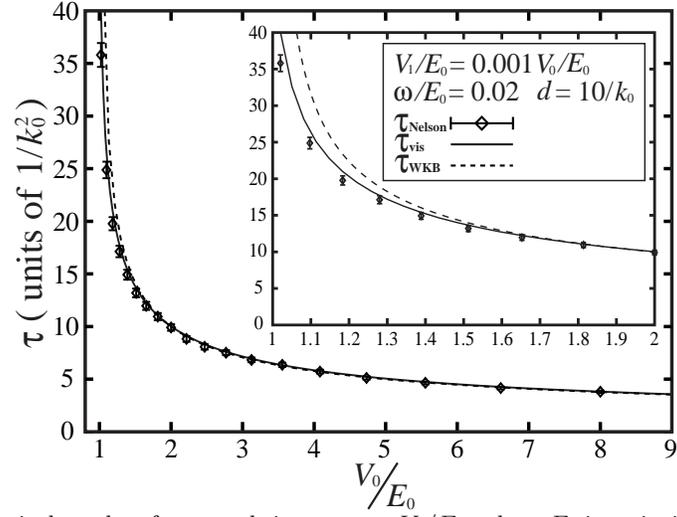}}
  \caption{Comparison of numerical results of traversal times versus $V_0/E_0$, where $E_0$ is an incident energy and $V_0$ is a potential height. The inset is a magnified part of small $V_{0}$.}
  \label{Fig.7}
\end{figure}

\end{document}